\documentstyle[a4,12pt,epsfig]{article}

\def\dfrac#1#2{{\displaystyle {#1 \over #2}}}

\def\simge{\mathrel{\rlap{\raise 0.511ex \hbox{$>$}}{\lower 0.511ex 
   \hbox{$\sim$}}}}
\def\simle{\mathrel{\rlap{\raise 0.511ex \hbox{$<$}}{\lower 0.511ex 
   \hbox{$\sim$}}}} 
\def\slash#1{\setbox0=\hbox{$#1$}\dimen0=\wd0 \setbox1=\hbox{/} \dimen1=\wd1 
  \ifdim\dimen0>\dimen1 \rlap{\hbox to \dimen0{\hfil/\hfil}} #1 \else     
  \rlap{\hbox to \dimen1{\hfil$#1$\hfil}} / \fi}
\newcommand{\be}{\begin{equation}}
\newcommand{\ee}{\end{equation}}
\newcommand{\bea}{\begin{eqnarray}}
\newcommand{\eea}{\end{eqnarray}}
   
\newcommand{\gev}{\,{\rm GeV}}   
\newcommand{\twb}{t\to Wb}   
\newcommand{\twzb}{t\to WZ\,b}   

\begin{document}

\begin{titlepage}
\begin{flushright}
CERN-TH/2000-170 \\
RM3-TH/00-12
\end{flushright}
\vskip 2.4cm
\begin{center}
{\Large \bf The $\twzb$ decay in the Standard Model: \\ \vspace{0.2cm}
A Critical Reanalysis}
\vskip1.3cm 
{\large\bf G.~Altarelli$^{a,b}$, L.~Conti$^b$ and V.~Lubicz$^b$}\\

\vspace{1.cm}
{\normalsize {\sl 
$^a$ Theory Division, CH-1211 Geneva 23, Switzerland \\
\vspace{.25cm}
$^b$ Dip. di Fisica, Univ. di Roma Tre and INFN,
Sezione di Roma III, \\
Via della Vasca Navale 84, I-00146 Rome, Italy\ .}}\\
\vskip1.cm
{\large\bf Abstract:\\[10pt]} \parbox[t]{\textwidth}{ 
We compute the $\twzb$ decay rate, in the Standard Model, at the leading order 
in perturbation theory, with special attention to the effects of the finite
widths of the $W$ and $Z$ bosons. These effects are extremely important, since 
the $\twzb$ decay occurs near its kinematical threshold. They increase the value
of the decay rate by orders of magnitude near threshold or allow it below the 
nominal threshold. We discuss a procedure to take into account the finite-width 
effects and compare the results with previous studies of this decay. Within the 
Standard Model, for a top quark mass in the range between 170 and 180 GeV, we 
find $BR(\twzb) \simeq 2 \times 10^{-6}$, which makes the observation at the LHC
very difficult if at all possible.}

\end{center}
\vspace*{1.cm} 

\end{titlepage}

\setcounter{footnote}{0}
\setcounter{equation}{0}

\section{Introduction
\label{sec:intro}}

The $\twzb$ process is a particularly interesting rare decay mode. In the 
Standard Model, all rare top decays are extremely suppressed with respect to 
the largely dominant $\twb$ decay. As possibly it could be observable at
the LHC, the $\twzb$ decay has already attracted considerable attention in the 
literature \cite{dnp}-\cite{jenkins}.

The most peculiar feature of the $\twzb$ decay is the fact that the process 
occurs near the kinematical threshold, $m_t \simeq m_W + m_Z + m_b$. As 
observed in ref.~\cite{mapa}, it is then crucial, in the theoretical 
evaluation of the decay rate, to take into account the finite-width effects 
for the $W$ and $Z$ bosons. For a top quark mass around approximately 175 GeV, 
these effects increase the value of the decay rate by some orders of 
magnitude. 

Even with the sizeable enhancement of the rate induced by finite-width 
effects, the branching ratio for $\twzb$ is predicted to be of the 
order of $10^{-6}$, which could be too small for the decay to be observed even 
at the LHC, where $10^{7}-10^{8}$ top quark pairs are expected to be produced 
per year. Therefore, the observation of this decay mode at the LHC could be a 
signal of physics beyond the Standard Model (see, for example, 
ref.~\cite{twohiggs,rizzo}). 

In this paper we first discuss the results for the  $\twzb$ decay rate 
computed at tree-level, in the Standard Model, by neglecting the effects of the
finite widths of the $W$ and $Z$ bosons, i.e. by treating these particles as 
stable particles in the final state. This quantity has also been computed in 
refs.~\cite{dnp}-\cite{jenkins} but the results of these studies are in 
disagreement. Our results, in the limit of vanishing widths, will be presented 
in sec.~\ref{sec:stables}. They agree with those obtained in 
ref.~\cite{jenkins}, but present numerical differences with respect to those 
of ref.~\cite{dnp} and, near threshold, of ref.~\cite{mapa}.

In sec.~\ref{sec:widths}, we compute the $\twzb$ decay rate by taking into 
account, as in ref.~\cite{mapa}, the finite-width effects of the $W$ and $Z$ 
bosons. We consider two different approaches. The first approach, which we 
will refer to as the {\it convolution method}, has been 
discussed, in a different context, in ref.~\cite{conv1}. It is based on 
a convolution of the $\twzb$ decay rate with two Breit-Wigner-like 
distributions for the invariant masses of the $W$ and $Z$ bosons. These 
distributions correspond to the imaginary part of the gauge boson propagators, 
and are centered around the physical values of the $W$ and $Z$ masses with 
spread controlled by the physical gauge boson widths. 

The second approach has been followed in ref.~\cite{mapa}. It consists in
considering the decay rate for the process $t \to W^* Z^* b$ followed by the 
decays of the virtual $W$ and $Z$ bosons, e.g. $W^* \to \mu\nu_\mu$ and $Z^* 
\to e^+ e^-$. The $\twzb$ decay rate is then obtained by dividing the rate 
computed for the full decay chain by the product of the $W\to \mu \nu_\mu$ and 
$Z\to e^+ e^-$ branching ratios. We will refer to this second approach as the 
{\it decay-chain method}. Incidentally, we note that one Feynman diagram 
entering at tree-level has not been included in the calculation of 
ref.~\cite{mapa}.

A difficulty inherent to the decay-chain method is that other initial processes,
besides the $\twzb$ decay of interest, may contribute to the decay chain. For
instance, when an electron pair is considered in the final state, a virtual 
photon, instead of the $Z$ boson, may be produced in the intermediate states.
Therefore, in this case, the direct connection between the full $t \to b \mu
\nu_\mu e^+ e^-$ process (total number of events) and the $\twzb$ decay 
(``signal") is lost. In order to suppress the contribution of the virtual 
photon, a kinematical cut on the invariant mass of the electron pair (requiring 
$m_{ee} \ge 0.8 \, m_Z$) has been introduced in ref.~\cite{mapa}. This cut is 
indeed effective and, in the selected region of the phase space, the $t \to b 
\mu\nu_\mu e^+ e^-$ process mostly proceeds through the initial $\twzb$ decay. 
However, with this requirement, the definition itself of the $\twzb$ decay rate 
becomes dependent on the specific choice of the cut. The purpose of this paper 
is to show that, by using the convolution method, a convenient definition of the
$\twzb$ decay rate can be provided, which is independent of any kinematical cut.
In this way, we find results for $\Gamma(\twzb)$ which differ by approximately 
a factor 3 with respect to those obtained following the procedure of 
ref.~\cite{mapa}. Of course, the full calculation of $t \to b \mu\nu_\mu e^+ 
e^-$, and the analysis of the kinematical cuts, can be relevant for the 
evaluation of the measurable signal plus background rate.

The calculation of the QCD radiative corrections goes beyond the scope of the
present paper. However their effect could be important and give rise to a
further reduction of the rate due to Sudakov suppression factors that tend to
become dominant near the end of phase space.

\section{The $\twzb$ decay rate in the limit of stable $W$ and $Z$ bosons
\label{sec:stables}}

\begin{figure}[t!]
\begin{center}
\begin{tabular}{c c c}
\hfill &\epsfxsize12.0cm\epsffile{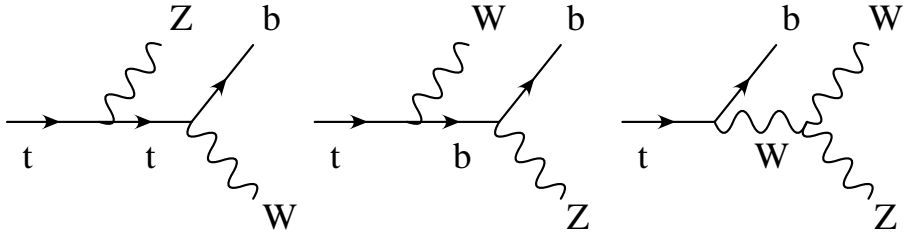}&\hfill  
\end{tabular}
\caption{\sl Feynman diagrams contributing at tree-level to the $\twzb$ decay.
\label{fig:fd_twzb}}
\end{center}
\end{figure}

Within the Standard Model, three Feynman diagrams contribute to the $\twzb$ 
decay at the leading order in perturbation theory. These diagrams are shown in 
Fig.~\ref{fig:fd_twzb} and represent the amplitudes with the final $Z$ boson 
radiated either by the initial top, or by the final $b$ or $W$ respectively.
%%%%%%%%%%%%%%%%%%%%%%%%%%%%%%%%%%%%%%%%%%%%%%%%%%%%%%%%%%%
%%%%%%%%%%%%%%%%%%%%%%%%%%%%%%%%%%%%%%%%%%%%%%%%%%%%%%%%%%%%%%%%%

%%%%%%%%%%%%%%%%%%%%%%%%%%%%%%%%%%%%%%%%%%%%%%%%%%%%%%%%%%%
\renewcommand{\arraystretch}{1.5}
\begin{table}
\begin{center}
\begin{tabular}{|ccccc|}
\hline \hline
$m_b$ & $m_W$ & $m_Z$ & $\sin^2\theta_W$ & $G_F$ \\ \hline
4.5 GeV & 80.3 GeV & 91.2 GeV & 0.23 & $1.166 \times 10^{-5}\;\gev^{-2}$ 
\\ \hline \hline
$\Gamma_W$ & $\Gamma_Z$ & $BR(W \to \mu \nu_\mu)$ & 
$BR(Z \to e^+e^-)$ & $BR(Z \to \nu \bar \nu)$ \\ \hline
2.06 GeV & 2.49 GeV & 0.102 & 0.03367 & 0.06667 \\ \hline \hline
\end{tabular}
\end{center}
\caption{\it Values of the numerical constants used in this paper  
\label{tab:constants}}
\end{table}
\renewcommand{\arraystretch}{1.0}
%%%%%%%%%%%%%%%%%%%%%%%%%%%%%%%%%%%%%%%%%%%%%%%%%%%%%%%%%%%
We have computed the decay rate for this process by considering first the 
final $W$ and $Z$ bosons as stable particles, i.e. neglecting their finite 
widths. In table \ref{tab:constants} we collect the values of the various 
quantities which have been used to obtain all the results presented in this 
paper. The integration over the final phase space of the three massive 
particles has been performed numerically. For the values of masses considered 
in this paper, the kinematical threshold for the process is at $m_t=176$ GeV. 
Note that the exact threshold value depends on $m_b$. At the parton level, 
$m_b$ is in principle scale dependent. A low scale appears justified here 
because of both the very limited phase space allowed to a virtual $b$ and the 
value of lightest $B$-meson mass.

In the limit of neglecting the finite widths of the $W$ and $Z$ bosons, the 
decay is forbidden below the threshold. For larger values of the top quark 
mass, we obtain the values of the branching ratio presented in table 
\ref{tab:br}.%
\footnote{Throughout this paper, we approximate the total decay width of the 
top quark with the partial rate $\Gamma(\twb)$, computed at the corresponding 
values of the top quark mass. Therefore, we define for instance the $\twzb$ 
branching ratio as:
$$ BR(\twzb) \equiv \frac{\Gamma(\twzb)}{\Gamma(\twb)} \, . $$}
The uncertainties on these results, coming from the numerical integrations, 
are estimated to be smaller than the last digit shown in the table. For values 
of the top quark mass around 180 GeV, we find that the branching ratio is 
extremely small, of the order of $10^{-8}$.
\renewcommand{\arraystretch}{1.3}
\begin{table}[t]
\begin{center}
\begin{tabular}{|c|ccc|} \hline \hline 
$m_{top}$ & 
$\begin{array}{cc} BR(\twzb) \\ {\mathrm Stable \ Particles} \end{array}$ &
$\begin{array}{cc} BR(\twzb) \\ {\mathrm Conv.\ Meth.} \end{array}$ &
$\dfrac{BR(t \to b \mu \nu_\mu \nu \bar\nu)}
{BR(W\to \mu \nu_\mu) BR(Z\to \nu \bar\nu)}$ \\ \hline
150 &	   -	          &$ 0.65(3)\cdot 10^{-6} $&$ 0.79(1)\cdot 10^{-6} $ \\
155 &	   -	          &$ 0.90(5)\cdot 10^{-6} $&$ 0.87(1)\cdot 10^{-6} $ \\
160 &	   -	          &$ 0.99(5)\cdot 10^{-6} $&$ 1.06(3)\cdot 10^{-6} $ \\
165 &	   -	          &$ 1.40(7)\cdot 10^{-6} $&$ 1.37(3)\cdot 10^{-6} $ \\
170 &	   -	          &$ 1.81(9)\cdot 10^{-6} $&$ 1.53(4)\cdot 10^{-6} $ \\
175 &	   -	          &$ 2.4 (1)\cdot 10^{-6} $&$ 1.96(5)\cdot 10^{-6} $ \\
180 &$ 0.07\cdot 10^{-6} $&$ 3.1 (2)\cdot 10^{-6} $&$ 2.76(8)\cdot 10^{-6} $ \\
185 &$ 0.63\cdot 10^{-6} $&$ 4.7 (2)\cdot 10^{-6} $&$ 4.0 (2)\cdot 10^{-6} $ \\
190 &$ 2.20\cdot 10^{-6} $&$ 7.4 (4)\cdot 10^{-6} $&$ 6.0 (3)\cdot 10^{-6} $ \\
195 &$ 5.37\cdot 10^{-6} $&$ 11.3(6)\cdot 10^{-6} $&$ 9.5 (5)\cdot 10^{-6} $ \\
200 &$ 10.7\cdot 10^{-6} $&$ 17.7(9)\cdot 10^{-6} $&$ 18  (2)\cdot 10^{-6} $ \\
\hline \hline
\end{tabular}
\caption{\it Values of the $\twzb$ branching ratio, as a function of the top 
quark mass, obtained in the limit of vanishing $Z$ and $W$ widths (stable
particles) and with the convolution method. In the last column, the estimate 
obtained with the decay-chain method from the $t \to b \mu \nu_\mu \nu \bar\nu$ 
decay is also shown for comparison.
\label{tab:br}}
\end{center}
\end{table}
\renewcommand{\arraystretch}{1.0}
%%%%%%%%%%%%%%%%%%%%%%%%%%%%%%%%%%%%%%%%%%%%%%%%%%%%%%%%%%%

The results obtained in this section agree with those of ref.~\cite{jenkins}. 
We also agree with the analytical expression of the tree-level Feynman 
amplitude squared published in that paper. On the other hand, we disagree with 
the numerical results of ref.~\cite{dnp} (we note that there is an apparent 
error in the integration limits over the phase space of the three massive 
particles (eq.(40) of that paper)).

We also found discrepancies, of the order of $50\%$, with respect to the 
results obtained in ref.~\cite{mapa} in what they call the {\it narrow-width 
approximation}. This procedure consists in de\-ri\-ving the values of the 
$\twzb$ decay rate starting from the calculation of $\Gamma(t \to b \mu\nu e^+
e^-)$ and constraining the invariant masses of the virtual $W$ and $Z$ bosons to
their physical values. In turn, these constraints are achieved by introducing 
{\it ad hoc} delta functions into the phase space integrals. We find that, in 
the stable particles limit, our results differ from those of ref.~\cite{mapa} 
by a factor 2 for $m_{top}=177$ GeV and by approximately 10\% for $m_{top} = 
185$ GeV.  
 
\section{The $\twzb$ decay rate including the finite-width effects
\label{sec:widths}}

In this section we discuss the calculation of the $\twzb$ decay rate by taking
into account the large finite-width effects of the $W$ and $Z$ bosons. We
present in detail the two approaches of the convolution and decay-chain method,
and critically compare the corresponding estimates for the decay rate.

\subsection{The convolution method
\label{sec:convo}}

Since the $W$ and $Z$ bosons are unstable particles with finite widths, their
production, in a physical process, can be approximately described as the 
production of real stable particles with invariant masses distributed 
according to a given distribution function. The central value and the width
of such a distribution are controlled by the physical mass and width of the 
unstable particle. From this point of view, for instance, the $\twzb$ decay, 
if occurs for values of the top quark mass smaller than the kinematical 
threshold, proceeds through the production of $W$ and $Z$ bosons with invariant
masses close, but smaller than, their physical values.

These observations suggest a very convenient way to take into account the 
finite-width effects in the $\twzb$ decay and, in general, in any process
with unstable particles produced in the final state. One computes the decay 
rate of the $\twzb$ channel as a function of generic values of the $W$ and $Z$ 
invariant masses. Let us denote this quantity as $\Gamma(k^2_W,k^2_Z)$, where 
$k_W^2$ and $k_Z^2$ are the virtualities of the $W$ and $Z$ bosons 
respectively. The $\twzb$ decay rate is then obtained by performing a 
convolution of $\Gamma(k^2_W,k^2_Z)$ with the two invariant-mass distribution
functions
\bea
\Gamma (\twzb) \!\!\! &=& \!\!\!
\int_0^{( m_t -m_b )^2} d k^2_Z \int_0^{( m_t -m_b -{\sqrt{k^2_Z}} )^2} d k^2_W 
\;\; \Gamma (k^2_W,k^2_Z) \times 
\nonumber \\ && \hspace{0.7cm} 
\times \;\; \rho(k^2_Z,M_Z,\Gamma_Z) \;\; \rho_(k^2_W,M_W,\Gamma_W)
\label{convolo}
\eea
The distribution $\rho(k^2,M,\Gamma)$ is related to the imaginary part of the 
gauge boson propagator: 
\be
\rho(k^2,M,\Gamma) = -\frac{1}{\pi} Im \left(\frac{1} { k^2-M^2+i M 
\Gamma(k^2)} \right) = \frac{1}{\pi} \frac{M \Gamma(k^2)} { \left(k^2-M^2
\right)^2 + M^2\, \Gamma(k^2)^2}
\label{rho}
\ee
where $M$ is the mass of the particle and $M\,\Gamma(k^2)$ the imaginary part 
of the vacuum po\-la\-ri\-za\-tion. In the calculation of $\Gamma(k^2)$ one 
may neglect, as a first approximation, the masses of the particles flowing in 
the loops. In this limit, by simple dimensional analysis, one gets:
\be
M \, \Gamma(k^2) = k^2 \frac{\Gamma}{M}
\label{partetrasv}
\ee
where $\Gamma\equiv\Gamma(M^2)$ is the particle width.%
\footnote{Note that the full $k^2$ dependence of $\Gamma(k^2)$ must be taken 
also into account in the numerator of eq.~(\ref{rho}). This dependence cancels 
out the singularity at $k^2=0$ appearing in the propagators and also in the sum 
over the polarization vectors of a particle with invariant mass $k^2$:
$$
\sum_{\lambda=1,2} \; \epsilon_{\lambda}^\mu(k) \epsilon_{\lambda}
^{\nu *}(k) = - g^{\mu \nu} + \frac{k^\mu k^\nu}{k^2} \, .
$$
In turn, this prescription for the sum over the polarization states is 
necessary to guarantee the positivity of the ``off-shell" decay rate 
$\Gamma (k^2_W,k^2_Z)$. We also note that in the calculation of 
$\Gamma (k^2_W,k^2_Z)$, since the invariant mass of the $W$-boson is allowed to 
vanish in eq.~(\ref{convolo}), we have modified the tree-level expression of 
the $W$-propagator as indicated in eq.~(\ref{gbprop}).}

In the limit of vanishing width, the distribution function (\ref{rho}) reduces
to the delta function $\delta(k^2-M^2)$, and the final particles are 
constrained on their mass shell. In this limit, eqs.~(\ref{convolo}) and 
(\ref{rho}) are just a consequence of the optical theorem and the Cutkosky 
rule to extract the imaginary part of the physical amplitude. In 
general, the procedure outlined above, which we refer to as the {\it 
convolution method}, represents the simplest way to take into account the 
finite-width effects in the evaluation of the decay rate, and it has been 
already considered in the literature in different contexts (see for example 
refs.~\cite{conv1,conv2}).

\begin{figure}[t!]
\begin{center}
\begin{tabular}{c c c}
\hfill &\epsfxsize11.0cm\epsffile{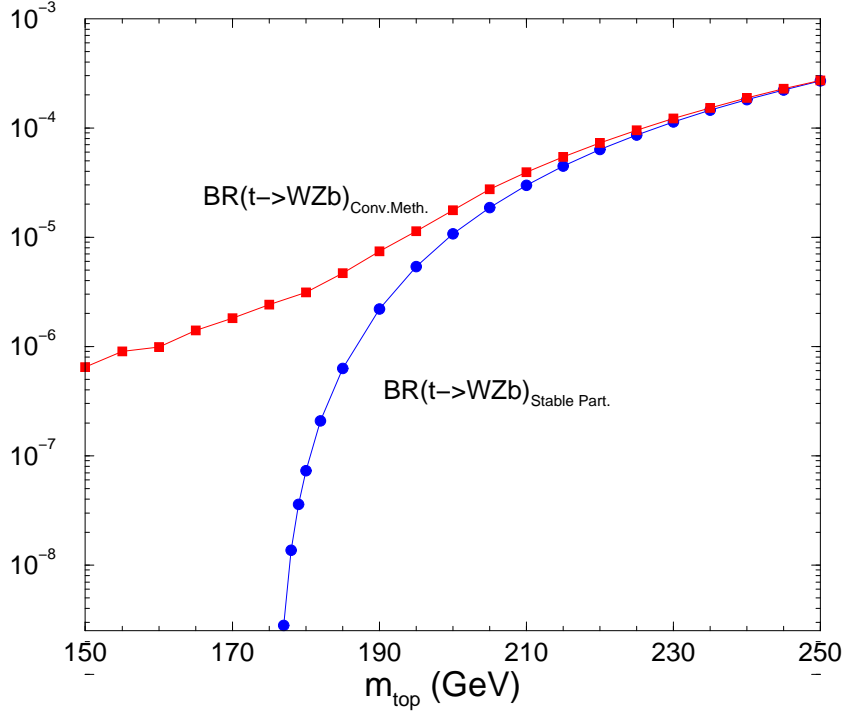}&\hfill  
\end{tabular}
\caption{\sl The $\twzb$ branching ratio, as a function of the top quark mass, 
as obtained in the stable particle limit and by taking into account the $W$ 
and $Z$ finite width effects with the convolution method.
\label{bratios}}
\end{center}
\end{figure}

The results obtained for the $\twzb$ branching ratio, by using the convolution
method, are presented in table \ref{tab:br}. These values are also shown in 
Fig.~\ref{bratios}, together with the corresponding values obtained in the 
stable-particles approximation, i.e. by neglecting the finite-width effects.
%%%%%%%%%%%%%%%%%%%%%%%%%%%%%%%%%%%%%%%%%%%%%%%%%%%%%%%%%%%
%%%%%%%%%%%%%%%%%%%%%%%%%%%%%%%%%%%%%%%%%%%%%%%%%%%%%%%%%%%%%%%%%
We see that, for a top quark mass around the kinematical threshold of 176 GeV, 
the effects of the finite widths increase the bran\-ching ratio by orders of 
magnitude, and allow the occurrence of the decay even below the threshold. 
However, despite the large enhancement induced by the widths, for the actual 
value of the top quark mass, the branching ratio is found to be of the order 
of $10^{-6}$, which is probably too small for the decay to be visible at the 
Tevatron Run II or even at the LHC.

The finite-width effects in the $\twzb$ decay rate have also been studied in
ref.~\cite{mapa}. By comparing our results with those obtained following the
procedure of ref.~\cite{mapa}, we find that the values of the decay rate are 
larger by approximately a factor 3. The reason of such a discrepancy will be 
discussed in detail in the next section.

\subsection{The decay-chain method
\label{sec:fourf}}

The procedure considered in ref.~\cite{mapa} to take into account the 
finite-width effects of the gauge bosons in the $\twzb$ decay is based on the 
study of a process which includes the decays of the unstable particles. For 
illustrative purposes, it is convenient to consider first the decay chain:
\be
t \to b W^* Z^* \; \Longrightarrow \; W^* \to \mu \nu_\mu, \;\; Z^* \to 
\nu \bar \nu
\ee
which differs, from the one studied in ref.~\cite{mapa}, for the presence of a
$\nu \bar \nu$ pair, rather than $e^+ e^-$, in the final state. 

In the calculation of the total rate $\Gamma(t \to b \mu \nu_\mu \nu \bar\nu)$, 
the finite-width effects of the intermediate $W$ and $Z$ bosons are then taken 
into account by correcting the tree-level expression of the gauge boson 
propagators. In the unitary gauge, by neglecting the $k^2$-dependence of the 
imaginary part of the vacuum polarization, the modified propagator can be 
written in the form~\cite{pestieau,apostol1}:
\be
\frac{i}{k^2 - (m - i\Gamma/2)^2} \left (- g^{\mu\nu} + \frac{k^\mu k^\nu}{ 
(m - i\Gamma/2)^2}
\right )
\label{gbprop}
\ee
where the correction to the longitudinal part of the propagator is necessary to 
respect gauge-invariance. In order to preserve unitarity at high energies, the 
absorbitive part of the triple gauge boson vertex should be also included in 
the amplitude~\cite{baur,apostol2}. However, in the case of interest, this 
correction is numerically negligible and for simplicity it has been omitted 
closely following the calculation of ref.~\cite{mapa}

The total $t \to b \mu \nu_\mu \nu \bar\nu$ rate is then used to compute the
ratio:
\be
\frac{\Gamma(t \to b \mu \nu_\mu \nu \bar\nu)}
{BR(W\to \mu \nu_\mu) BR(Z\to \nu \bar\nu)}
\label{usuale}
\ee
which is assumed to be an estimate of the $\twzb$ decay rate.

\begin{figure}[t!]
\begin{center}
\begin{tabular}{c c c}
%\hfill &\epsfxsize11.0cm\epsffile{diamiss.eps}&\hfill  
\hfill &\epsfxsize6.0cm\epsffile{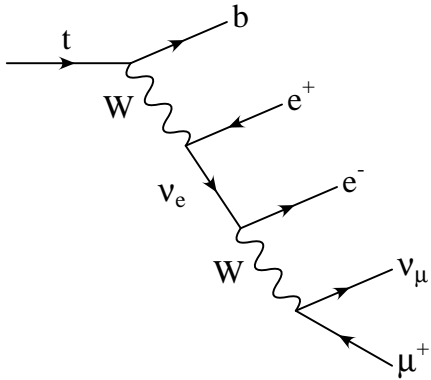}&\hfill  
\end{tabular}
\caption{\sl A Feynman diagram, contributing at tree-level to the $t \to b 
\mu \nu_\mu e^+e^-$ decay, which has not been considered in the calculation of 
ref.~\protect{\cite{mapa}}.
\label{fig:missed}}
\end{center}
\end{figure}

A crucial requirement, for the above procedure to be consistent, is that the
full decay chain, which is considered to estimate $\Gamma(\twzb)$, can, at
least dominantly, only proceed through the initial $\twzb$ decay. Otherwise, 
any direct connection with this process is obviously lost. This requirement is
only partially satisfied in the case of the $t \to b \mu \nu_\mu \nu \bar\nu$
decay. Three Feynman diagrams, describing at tree-level the $t \to b \mu \nu_\mu
\nu \bar \nu$ decay (with $\nu \neq \nu_\mu$), are those shown in 
Fig.~\ref{fig:fd_twzb}, modified to account for the $W$ and $Z$ bosons decays 
into the $\mu \nu_\mu$ and $\nu \bar \nu$ pairs respectively. These diagrams
describe the processes which proceed through the initial $\twzb$ decay. Besides 
these graphs, however, other three Feynman diagrams also contribute to the 
decay. They are non-resonant diagrams, similar to that of Fig.~\ref{fig:missed},
in which one of the intermediate gauge bosons is produced away from its 
mass-shell. In table \ref{tab:br} we present the values of the branching ratio 
obtained by using eq.~(\ref{usuale}). By comparing these values with the $\twzb$
branching ratio evaluated by using the convolution method we find that the 
results are in reasonable agreement. It should be noticed, however, that this 
agreement is also a consequence of an accidental cancellation between the square
of the non-resonant diagrams and the interference among these diagrams and those
proceeding through the initial $\twzb$ decay. Ideally, one would like to 
consider in the calculation of the decay rate only these three latter diagrams. 
However, such a prescription does not lead, in general, to a gauge-invariant 
result. 

The decay-chain method has been considered in ref.~\cite{mapa} to account for 
the finite-width effects of the $W$ and $Z$ bosons in the $\twzb$ decay. 
In that case, the decay rate has been evaluated from the ratio:
\be
\frac{\Gamma(t \to b \mu \nu_\mu e^+e^-)}{BR(W\to \mu \nu_\mu) BR(Z\to e^+e^-)}
\label{errore}
\ee
which differs from eq.~(\ref{usuale}) for the choice of the final state. The 
calculation of $\Gamma(t \to b \mu \nu_\mu e^+e^-)$ in eq.~(\ref{errore})
requires to take into account a set of 10 Feynman diagrams. The first three 
diagrams are those proceeding through the initial $\twzb$ decay. This is the 
only contribution which is directly related to $\Gamma(\twzb)$. Other three 
diagrams involve the radiation of a $Z$ or a $W$ boson from one of the decay 
product of the first $W$. These are non-resonant diagrams in which the 
intermediate $W$ is produced away from its mass-shell. The last four diagrams 
are those in which the intermediate $Z$ boson is replaced by a virtual photon. 
In the region of small invariant masses of the final electron pair, these are 
the diagrams which give the dominant contribution to the decay rate.

\begin{figure}[t!]
\begin{center}
\begin{tabular}{c c c}
\hfill &\epsfxsize11.0cm\epsffile{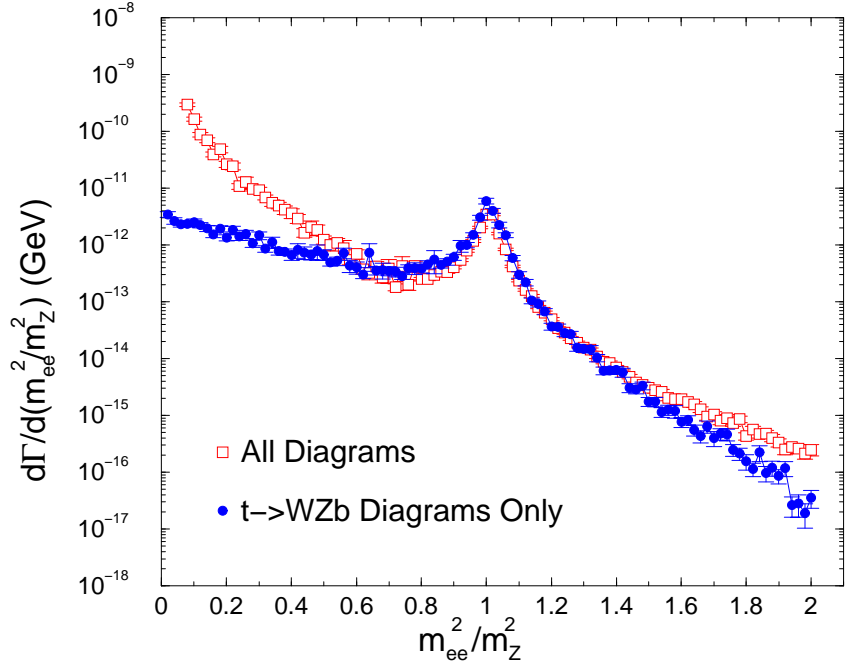}&\hfill  
\end{tabular}
\caption{\sl The differential rate $d\Gamma/d(m^2_{ee}/m_Z^2)$, for the $t\to b
 \mu \nu_\mu e^+ e^-$ decay, computed by considering all Feynman diagrams 
contributing to the process or only the three relevant diagrams which proceeds
through the initial $\twzb$ decay. The top quark mass is fixed to $m_t=175$ 
GeV.
\label{fig:gammaz}}
\end{center}
\end{figure}

Two observations are worth at this point. 

i) In evaluating the Feynman diagrams, the authors of ref.~\cite{mapa} have 
missed the contribution of the non-resonant diagram shown in 
Fig.~\ref{fig:missed}.
%%%%%%%%%%%%%%%%%%%%%%%%%%%%%%%%%%%%%%%%%%%%%%%%%%%%%%%%%%%
%%%%%%%%%%%%%%%%%%%%%%%%%%%%%%%%%%%%%%%%%%%%%%%%%%%%%%%%%%%
Although numerically small, in the unitary gauge, this contribution is necessary
to guarantee a gauge-invariant final result for the decay rate. In addition, in 
the Standard Model, the $t \to b \mu \nu_\mu e^+e^-$ decay also receives a 
tree-level contribution from a set of four Feynman diagrams in which the virtual
$Z$ is replaced by a Higgs boson. Also these diagrams have not been taken into 
account in ref.~\cite{mapa}. We find that, numerically, the contributions of 
these diagrams is negligible for values of the Higgs mass larger than 
approximately 100 GeV.

ii) In the theoretical evaluation of the $t\to b \mu\nu_\mu e^+e^-$ decay rate, 
all Feynman diagrams contributing to the process, at a given order in 
perturbation theory, must be taken into account. However, as discussed before, 
because of the contribution of the virtual photon, with the prescription in 
eq.~(\ref{errore}) any direct connection between the $t \to b \mu \nu_\mu e^+
e^-$ and the $\twzb$ decays is lost. To show it clearly, we plot in 
Fig.~\ref{fig:gammaz} the differential $t \to b \mu \nu_\mu e^+e^-$ decay rate, 
as a function of the invariant mass of the electron pair, as obtained either by 
including in the calculation all Feynman diagrams contributing to the process 
or only the three diagrams which involve the initial $\twzb$ decay.
%%%%%%%%%%%%%%%%%%%%%%%%%%%%%%%%%%%%%%%%%%%%%%%%%%%%%%%%%%%
%%%%%%%%%%%%%%%%%%%%%%%%%%%%%%%%%%%%%%%%%%%%%%%%%%%%%%%%%%%%%%%%%
Although the latter do not form a gauge-invariant set of Feynman diagrams, the
comparison is instructive. The numerical discrepancy, which is seen between the
result of the full calculation and the genuine $\twzb$ contribution, it is 
mainly determined by the contribution of the virtual photon in the region of 
small values of the $e^+e^-$ invariant mass. In order to reduce this 
contribution, the authors of ref.~\cite{mapa} have introduced a cut on the 
values of this mass, requiring $m_{ee} \ge 0.8 \, m_Z$. As can be seen from 
Fig.~\ref{fig:gammaz}, this cut is indeed effective to that purpose, and in the
selected kinematical region the $t \to b \mu \nu_\mu e^+e^-$ decay mainly 
proceeds through the initial $\twzb$ decay. However, with this prescription, 
the definition itself of the $\twzb$ decay rate becomes dependent on the 
specific choice of the cut. 

As we have shown before, a convenient definition of the $\twzb$ decay rate, 
which is independent of any kinematical cut, is provided by the convolution 
method. The results obtained in this way for the $\twzb$ decay rate are roughly 
3 times larger than those obtained in ref.~\cite{mapa}, the main reason being 
the introduction of the kinematical cut in the definition of the decay rate. On 
the other hand, we emphasize that the calculation of ref.~\cite{mapa} is of 
interest for the evaluation of the signal and the relevant background rates in 
the experimental study of the $\twzb$ decay when the $b \mu \nu_\mu e^+e^-$ is 
considered as a final state.

\section{Conclusion
\label{sec:conclu}}
In this paper we have computed the $\twzb$ decay rate, at the leading 
perturbative order in the Standard Model, by including in the calculation the 
crucial effects of the $W$ and $Z$ widths. We finds that these effects 
increase the total decay rate by orders of magnitude, for a top quark mass 
of approximately 176 GeV, and allow the decay below the nominal threshold.

In the limit in which the finite-width effects are neglected, the results of 
the previous studies of the $\twzb$ decay \cite{dnp}-\cite{jenkins} are in 
disagreement among each other. We have repeated the calculation in this limit, 
and obtained predictions in agreement with those of ref.~\cite{jenkins}, but
with numerical differences with respect to those of ref.~\cite{dnp} and, near 
threshold, of ref.~\cite{mapa}. 

We have also shown that a practical definition of the $\twzb$ decay rate can
be provided even in the presence of large finite-width effects of the $W$ and
$Z$ bosons. This definition is based on a convolution of the decay rate, 
computed in the limit of stable particles, with two distribution functions for 
the invariant masses of the gauge bosons. The distributions are related to the 
imaginary part of the gauge boson propagator, and are centered around the 
physical value of the $W$ and $Z$ masses with spread controlled by the particle 
widths. The results obtained in this way for $\Gamma(\twzb)$ differ, by roughly 
a factor 3, from those obtained following the procedure of ref.~\cite{mapa}. 
The main reason of such a discrepancy is the introduction of a kinematical cut 
which is included, in ref.~\cite{mapa}, in the definition of the decay rate.

As a final estimate of the $\twzb$ branching ratio, evaluated by taking into
account the finite-width effects, we get: 
\be
BR(\twzb) \simeq 2 \times 10^{-6}
\label{eq:final}
\ee
for a top quark mass in the range between 170 and 180 GeV. Eq.~(\ref{eq:final})
indicates that, within the Standard Model, the branching ratio is quite small 
so that the observation of the decay at the LHC is extremely difficult, if at 
all possible. Alternatively, the observation of the $\twzb$ decay at the LHC 
with a larger rate would signal the presence of new physics.

\section*{Acknowledgments}
We are grateful to B.~Mele and S.~Petrarca for interesting discussions on the 
subject of this paper. We also thank A.~Pukhov for his precious suggestions 
concerning the use of the program CompHEP.

\end{document}